\documentclass[12pt]{article}
\def\Hline{\noalign{\hrule height .5mm}}
\def\Vline{\vrule width .5mm}
\begin{document}
\begin{center}
{\Large \bf Exactly Solvable Chaos and 
Addition Theorems of Elliptic Functions}\\   
\vskip.25in
{ Ken Umeno}\footnote{Present Address: Communications Research Laboratory,
Ministry of Posts and Telecommunications, 4-2-1, Nukui-Kitamachi, Koganei,
Tokyo 184-8795,Japan}
 
{\it Frontier Research Program,\\
 The Institute of Physical 
  and Chemical Research (RIKEN),\\ 
2-1 Hirosawa, Wako, Saitama 351-01, Japan \\} 
(December 24, 1998)
\vskip.25in
\end{center}

\begin{abstract}
A unified view is given to recent developments about
a systematic method of  constructing rational mappings
as ergodic transformations with non-uniform 
invariant
measures on the unit interval \(\mbox{\boldmath$I$}=\left[0,1\right]\).
 All of the rational ergodic mappings  of \(\mbox{\boldmath$I$}\) 
with explicit non-uniform invariant densities  can be 
 obtained by addition theorems of elliptic functions. 
It is shown here that the class of the rational ergodic 
mappings \(\mbox{\boldmath$I$}\to \mbox{\boldmath$I$}\) are essentially same as 
the permutable rational functions obtained by J. F. Ritt.
\end{abstract}
  
\clearpage 
\setcounter{equation}{0} 
\section{Introduction}
 Qualitative characterizing  discrete-time dynamical systems have been 
recently gaining an  attention. In continuous-time dynamical systems given
by  ordinary differential equations (ODE), the notion of integrability
can serve as the sharp characteristic for general ODEs such that integrable
ODEs are solvable in the sense  that  exact solutions are analytically
obtained and non-integrable ODEs such as the three-body problems are not
solvable and they show, in general, chaotic 
behavior. In contrast with ODE cases,
 the notion of integrability itself is still vague in discrete-time systems,
that is,  it is not so sharp  as the case of ODEs. Quite recently,
Hietarinta and  Viallet 
show examples of discrete-time dynamical systems which pass 
the singularity confinement test (discrete-time version of Painl\'eve test), 
but which nevertheless show chaotic behavior \cite{hietarinta}.  
Thus, an issue  is  now focused on searching a characteristic distinguishing
chaotic 
discrete-time systems from regular discrete-systems (which have been thought
as {\it integrable} systems.    
The purpose of the present paper is to present a class of  
discrete-time dynamical systems which have exact solutions, but which
nevertheless are shown to be ergodic (thus, chaotic). Such dynamical systems
(we call {\it exactly solvable chaos}) can be 
 constructed in 
a systematic way, by utilizing addition theorems of elliptic functions. 
The main claim here is that solvable discrete-time dynamical systems with
chaotic properties do not belong to  exceptional classes but they are as  
{\it ubiquitous as multiplication formulas of elliptic functions}, which
have  interesting implications to the issue.  
on the issue. 
\section{Construction}
Let us consider an elliptic function \(s(x)\) whose inverse function is
defined by the following elliptic integral: 
\begin{equation}
\label{eq:elliptic_f}
  s^{-1}(x)=\int_{0}^{x}\frac{du}{\sqrt{a_{0}+a_{2}u^{2}+a_{4}u^{4}
+a_{6}u^{6}}}
=\int_{0}^{x^{2}}\frac{dv}{2\sqrt{v(a_{0}+a_{2}v+a_{4}v^{2}+a_{6}v^{3})}},
\end{equation}
 where 
%\(K\) is an elliptic integral  
%\begin{equation}
%    K=\int_{0}^{1}\frac{du}
%{\sqrt{a_{0}+a_{1}u+a_{2}u^{2}+a^{3}u^{3}+a^{4}u^{4}}},
%\end{equation}
 \(0\leq x \leq 1\)
 and \(a_{0},\cdots a_{4}\) are real-valued constants, which 
satisfy the following condition:
\begin{equation}
\label{eq:condition1}
\begin{array}{l}
a_{0}+a_{2}u^{2}+a_{4}u^{4}+a_{6}u^{6}>0\quad \mbox{for}\quad 0<u<1\\
a_{0}+a_{2}+a_{4}+a_{6}=0, a_{0}> 0.
\end{array}
\end{equation}  
This elliptic function can be  also considered as the solution \(q(t)\) of
a Hamiltonian system the motion  with 
a Hamiltonian 
\begin{equation}
  H=\frac{1}{2}p^{2}+V(q)=0, 
\end{equation} 
where the potential function \(V(q)\) is an even polynomial function of the
form 
\(V(q)=-(a_{0}+a_{2}q^{2}+a_{4}q^{4}+a_{6}q^{6})\) satisfying the conditions 
\begin{equation}
   V(0)=0,\quad V(1)=0, \quad V(q)<0\quad \mbox{for}\quad 0<q<1.
\end{equation} 
Thus on this condition about \(a_{0},\cdots,a_{6}\) in Eq.
(\ref{eq:condition1}), 
this elliptic function \(s(x)\) has a real period \(4K\), where \(K\) 
is defined by the elliptic integral
\begin{equation}
K=\int_{0}^{1}\frac{du}{\sqrt{a_{0}+a_{2}u^{2}+a_{4}u^{4}
+a_{6}u^{6}}}.
\end{equation}
We have the  relations
\begin{equation}
0\leq s(x)\leq 1\quad \mbox{for}\quad 0\leq x \leq K,
\end{equation}
and 
\begin{equation} 
 s(0)=0, \quad s(K)=1,\quad s(2K)=0,\quad s(3K)=-1,\quad s(4K)=0.
\end{equation}
Since the equality 
\begin{equation}
  s^{-1}(-x)=\int_{0}^{-x}\frac{du}
{\sqrt{a_{0}+a_{2}u^{2}+a^{4}u^{4}+a^{6}u^{6}}}=-s^{-1}(x)\equiv y
\end{equation}
holds, then we have 
\begin{equation}
  s(y)=-x,\quad s(-y)=x
\end{equation} 
which means that \(s^{2}(y)\) is an {\it even function} 
as \(s^{2}(y)=s^{2}(-y)\) and has a real period \(2K\).  
Furthermore, 
since any   elliptic function  has an {\it algebraic addition theorem}
\cite{addition}, 
there exists a polynomial function \(F\) in three variables such that 
the relation        
\begin{equation}
   F(s^{2}(x_{1}+x_{2}),s^{2}(x_{1}),s^{2}(x_{2}))=0
\end{equation} 
holds.
As a special case of the addition theorems,  
there exists a polynomial function \(G\) in two variables such that 
\begin{equation}
   G(s^{2}(px),s^{2}(x))=0, 
\end{equation} 
where \(p\) is a positive integer greater than the unity. 
This means that  
we have an algebraic mapping as 
\begin{equation}
\label{eq:addition_p}
   s^{2}(px)=f(s^{2}(x)),
\end{equation}
 where \(f\) is an algebraic function.
If we set 
\begin{equation}
X_{n+1}\equiv s^{2}(px),\quad X_{n}\equiv s^{2}(x),
\end{equation}
 we have a discrete-time dynamical system  
\begin{equation}
   X_{n+1}=f(X_{n}).
\end{equation}
 on the unit interval \(\mbox{\boldmath$I$}=[0,1]\).
It is easy to check that there exist \(p\) pre-images 
\begin{equation}
 0< x_{m}=s^{2}[\frac{2mK+s^{-1}(\sqrt{y})}{p}]<1,\quad 0\leq m\leq p-1
\end{equation}
satisfying \(y=f(x_{m})\) for \(0<y<1\). 
 This dynamical system \(X_{n+1}=f(X_{n})\)
 has the following remarkable property:
\newtheorem{th1}{Theorem}
\begin{th1}
\label{th1:density}
A dynamical system \(X_{n+1}=f(X_{n})\) defined as 
the addition theorem (\ref{eq:addition_p}) of an elliptic function 
\(s^{2}(x)\)  is ergodic with respect to  an invariant measure \(\mu(dx)\)
which is absolutely continuous with respect to the Lebesgue measure and 
their density function \(\rho(x)\) is given by the formula
\begin{equation}
\label{eq:density}
\rho(x)=\frac{1}
{2K\sqrt{x(a_{0}+a_{2}x+a_{4}x^{2}+a_{6}x^{3})}}.
\end{equation}
\end{th1}
({\bf Proof of Theorem \ref{th1:density}})\\

Let us consider the diffeomorphisms of \(\mbox{\boldmath$I$}=[0,1]\) 
into itself given by 
\begin{equation}
 0\leq  \phi(x)=\frac{1}{K}s^{-1}(\sqrt{x})\leq 1\quad \mbox{for}\quad
  0\leq x\leq 1.
\end{equation}
Using the relations
\begin{equation}
\begin{array}{l}
  s^{2}(K\cdot p\theta)=f[s^{2}(K\theta)]\quad \mbox{for}\quad \theta\in
[0,\frac{1}{p}],\\
  s^{2}[K(2-p\theta)]=f[s^{2}(K\theta)]\quad \mbox{for}\quad\theta\in
[\frac{1}{p},\frac{2}{p}],\\
\cdots,\\
s^{2}[K(-(-1)^{i}i+\frac{1-(-1)^{i}}{2}+(-1)^{i}p\theta)]
=f[s^{2}(K\theta)]\quad \mbox{for}\quad\theta\in
[\frac{i}{p},\frac{i+1}{p}],\\
\cdots,\\
 s^{2}[K((-1)^{p}(p-1)+\frac{1+(-1)^{p}}{2}+(-1)^{p-1}p\theta)]
=f[s^{2}(K\theta)]\quad \mbox{for}\quad \theta\in
[\frac{p-1}{p},1]
\end{array}
\end{equation}
by the fact that  \(s^{2}(x)\) is an even function and it has 
a real period \(2K\), we can derive the piecewise-linear map 
\(\tilde f(x)= \phi \circ f\circ \phi^{-1}(x)\)
 on  \(\mbox{\boldmath$I$}=[0,1]\) as 
\begin{equation}
\label{eq:piecewise}
\begin{array}{l}
\tilde f(x)=px\quad \mbox{for}\quad x\in[0,\frac{1}{p}]\\
\cdots,\\
\tilde f(x)=-(-1)^{i}i+\frac{1-(-1)^{i}}{2}+(-1)^{i}px \quad 
   x\in [\frac{i}{p},\frac{i+1}{p}]\\
\cdots,\\
\tilde f(x)=-(-1)^{p-1}(p-1)+\frac{1-(-1)^{p-1}}{2}+(-1)^{p-1}px\quad 
 x\in [\frac{p-1}{p},1]\\.
\end{array}
\end{equation}
It is noted here that when \(p=2\), 
\(\tilde f(x)\) is the {\it tent} map, given 
by 
\begin{equation}
\begin{array}{l}
\tilde f(x)=2x\quad x\in [0,\frac{1}{2}]\\
\tilde f(x)=2-2x\quad x\in [\frac{1}{2},1]. 
\end{array}
\end{equation}
Clearly, the map \(\tilde f(x)\) in Eq. (\ref{eq:piecewise}) is ergodic with 
respect to 
  the Lebesgue measure on \(\mbox{\boldmath$I$}=[0,1]\). 
Thus,  \(f\) is also ergodic with respect to the measure 
\begin{equation}
  \mu(dx)=\frac{d\phi(x)}{dx}dx=\frac{1}{2K}
\frac{dx}{\sqrt{x(a_{0}+a_{2}x+a_{4}x^{2}+a_{6}x^{3})}}.
\end{equation}
The measure \(\mu(dx)\) is absolutely continuous with respect to 
 the Lebesgue measure, by which we can define the algebraic density 
function \(\rho(x)\) in Eq.(\ref{eq:density}).\\
({\bf End of proof})\\

A simple corollary of the theorem is the following: According to the Pesin
identity which can be applied to the dynamical systems with absolutely
continuous measure\cite{pesin}, 
the Lyapunov characteristic exponent \(\Lambda(\mu)\) for the measure 
of the map \(X_{n+1}=f(X_{n})\) is equivalent to 
the Kolmogorov-Sinai entropy \(h(\mu)=\log p\), namely:
\begin{equation}
  \Lambda(\mu)=\int_{0}^{1}\log|\frac{df}{dx}|\cdot\rho(x)dx=h(\mu)=\log p.
\end{equation} 
 
In this sense, discrete-time  dynamical systems \(X_{n+1}=f(X_{n})\) 
constitute a typical class of {\it chaotic} dynamical 
systems with a special property 
that their invariant density functions are explicitly given. 
Hence, we 
call this class of chaotic dynamical systems {\it exactly solvable chaos}
\cite{ku3}. 
We note here that dynamical systems of exactly solvable chaos can be  
{\it good} pseudo random-number generators\cite{ku4}, especially for 
Monte Carlo simulations from the exact ergodicity
\begin{equation}
  \lim_{N\rightarrow \infty}
\sum_{n=0}^{N-1}\frac{1}{N}Q(x_{n}) = 
\int_{0}^{1}Q(x)\rho(x)dx.
\end{equation} 
We remark here that we can generate 
infinitely concrete chaotic dynamical systems with the unique 
invariant measure (\ref{eq:density}) from addition theorems 
\(s^{2}(px)=f[s^{2}(x)]\) of elliptic functions \(s(x)\) 
for \(p=2,3,\cdots\).
     
\section{Examples of exactly solvable chaos}
Historically, 
the most simple example of exactly solvable chaos 
is the logistic map \(Y=4X(1-X)\equiv f_{0}(X)\) on 
\(\mbox{\boldmath$I$}=\left[0,1\right]\)  given 
by Ulam and von Neumann in the late 1940's\cite{uf}. 

Ulam and von Neumann show that 
\(X_{n+1}=4X_{n}(1-X_{n})\) is ergodic with respect to 
 an invariant probability measure \(\mu(dx)=\frac{dx}{\pi\sqrt{x(1-x)}}\).
The Ulam-von Neumann map can be seen as a special case of 
exactly solvable chaos of the above theorem 1
when we consider \(s(x)\) defined by 
\begin{equation}
\label{eq:ulam_eq}
  s^{-1}(x)=\sin^{-1}(x)=\int_{0}^{x}\frac{du}{\sqrt{1-u^{2}}}.
\end{equation} 
This corresponds to the case that  
\begin{equation}
a_{0}=1,\quad a_{2}=-1,\quad a_{4}=a_{6}=0
\end{equation}
in Eq. (\ref{eq:elliptic_f}).
Thus, we have the relation \(s(x)=\mbox{sin} (x)\) which   
admits the duplication formula \(\sin^{2}(2x)=4\sin^{2}(x)(1-\sin^{2}(x))
\equiv f(\sin^{2}(x))\) (\(p=2\) in Eq.(\ref{eq:addition_p})) 
gives the logistic map. Clearly, in this case, \(K=\frac{\pi}{2}\).
Thus, using Theorem \ref{th1:density},
we have calculate the density function \(\rho (x)\) as 
\begin{equation}
\rho(x)=\frac{1}{\pi\sqrt{x(1-x)}}.  
\end{equation}
When \(p=3\), we have {\it the cubic map} as \(X_{n+1}=X(3-4X)^{2}\) from 
the triplication formula 
\begin{equation}
\sin^{2}(3x)=f[\sin^{2}(x)]=\sin^{2}(x)[3-4\sin^{2}(x)]^{2}
\end{equation}
The cubic map can be regarded as a special class of 
Chebyshev maps obtained by using more general addition formulas 
as \(\sin^{2}(px)=f(\sin^{2}(px))\)\cite{adler}.

The alternative attempts to  generalize the 
Ulam-von Neumann map within a set of rational functions was made by 
Katsura and Fukuda in 1985\cite{kf}.  The Katsura-Fukuda map is 
given by   
\begin{equation}
\label{eq:katsura}
Y=\frac{4X(1-X)(1-lX)}{(1-lX^{2})^{2}}\equiv f_{l}(X)  
\end{equation} 
 for \(0\leq l < 1\). Clearly, 
the Ulam-von Neumann map can be regarded as  a special 
case of  
Katsura-Fukuda maps. 
In this case, the 
corresponding elliptic function \(s(x)\) is Jacobi \(sn\) function 
whose inverse function is defined by 
\begin{equation}
  s^{-1}(x)=sn^{-1}(x;\sqrt{l})
=\int_{0}^{x}\frac{du}{\sqrt{(1-u^{2})(1-lu^{2})}},
\end{equation}
where \(\sqrt{l}\) corresponds to the modulus of Jacobi elliptic functions.
This corresponds to the case that 
\begin{equation}
  a_{0}=1,\quad a_{2}=-l+1,\quad, a_{4}=l,\quad a_{6}=0
\end{equation}
in Eq. (\ref{eq:elliptic_f}).
Katsura and Fukuda show \cite{kf} that the map (\ref{eq:katsura}) 
has exact solutions \(X_{n}=sn^{2}(2^{n}\theta;\sqrt{l})\). 
Using the idea of theorem 1, the author  shows  \cite{ku3} 
that the Katsura-Fukuda maps (\ref{eq:katsura}) are also ergodic with
respect to an  invariant 
measures which can be written explicitly as 
\begin{equation}
\label{eq:kfi}
\mu(dx)=\rho(x)dx= \frac{dx}{2K(l)\sqrt{x(1-x)(1-lx)}},
\end{equation} 
where \(K(l)\) is the elliptic integral of the first kind  given by 
\(
  K(l) = \int_{0}^{1} \frac{du}{\sqrt{(1-u^{2})(1-lu^{2})}}
\).
This can be  easily checked  
 using the duplication 
formula\cite{ww} of the Jacobi \(sn\) elliptic function  
\begin{equation}
sn(2u;\sqrt{l})=\frac{2sn(u;\sqrt{l})
\sqrt{(1-sn^{2}(u;\sqrt{l}))(1-lsn^{2}(u;\sqrt{l}))}}{(1-lsn^{4}(u;\sqrt{l}))}.
\end{equation} 

Recently, 
the more general classes of exactly solvable chaos are derived from
elliptic functions
\(s(x)\) whose inverse functions are defined by 
\begin{equation}
\label{eq:slm}
 s^{-1}(x)=\int_{0}^{x}\frac{du}{\sqrt{
(1-u^{2})(1-lu^{2})(1-mu^{2})}},
\end{equation}
where the  parameters \(l\) and \(m\) are arbitrary real numbers 
satisfying the condition 
 \(-\infty <m\leq l<1\) \cite{ku3}.
This corresponds to the case that 
\begin{equation}
  a_{0}=1,\quad a_{2}=-(l+m+1),\quad a_{4}=lm+l+m,\quad a_{6}=-lm
\end{equation}
 in Eq. (\ref{eq:elliptic_f}). 
The associated mapping generated by the duplication theorem 
\(s^{2}(2x)=f^{(2)}_{l,m}(s^{2}(x))\) is given by  
 the following rational  transformations  
\begin{equation}
\label{eq:formula1}
f^{(2)}_{l,m}(X)=\frac{4X(1-X)(1-lX)(1-mX)}{1+AX^{2}+BX^{3}+CX^{4}}\in I,
\end{equation}
  where 
\(A=-2(l+m+lm), 
 B=8 lm,
C= l^{2}+m^{2}-2lm-2l^{2}m-2lm^{2}+l^{2}m^{2}\), and 
\(X\in I\)\cite{ku3}. 
Clearly, this mapping is a generalized version
 of Ulam-von Neumann map and Katsuda-Fukuda map. 
This two-parameter family of the dynamical systems 
\(x_{n+1}=f^{(2)}_{l,m}(x_{n})\)  (\ref{eq:formula1}) are also ergodic with
respect to an invariant measures given by 
\begin{equation}
\label{eq:density_lm}
  \mu(dx)=\rho(x)dx=
  \frac{dx}{2K(l,m)\sqrt{x(1-x)(1-lx)(1-mx)}},
\end{equation}
 where \(K\) is given by the   
 integrals  
\begin{equation}
  K(l,m)=\int_{0}^{1}\frac{du}{\sqrt{(1-u^{2})(1-lu^{2})
   (1-mu^{2})}}.
\end{equation}
We can check this fact by directly computing  the duplication formula
of \(s(x)\) whose inverse function is defined in Eq. (\ref{eq:slm}). 
Let us represent 
\(s(x)\) 
in terms of the Weierstrass elliptic 
functions\cite{ww1}. 

The Weierstrass elliptic function \(\wp(u)\) of 
\(u\in \mbox{\boldmath$C$}\) is defined by 
\begin{equation}
  \wp(u)=\frac{1}{u^{2}}+{\sum_{j,k}}^{'}\{
\frac{1}{(u-2j\omega_{1}-2k\omega_{2})^{2}}-
\frac{1}{(2j\omega_{1}+2k\omega_{2})^{2}}\},
\end{equation}
  where the symbol \(\sum'\) means that the summation is made over 
all combinations of integers \(j\) and \(k\), except for the combination 
\(j=k=0\), and  \(2\omega_{1}\) and \(2\omega_{2}\) are  periods of 
the function \(\wp(u)\)\cite{ww}. The Weierstrass elliptic function 
\(\wp(u)\) 
satisfies the differential equation
\begin{equation}
\label{eq:we}  
  (\frac{d\wp (x)}{dx})^{2}=4\wp^{3}(x)-g_{2}\wp(x)-g_{3},
\end{equation}
 with the invariants
\begin{equation}   
  g_{2}(\omega_{1},\omega_{2}) =
 60{\sum_{j,k}^{'}}\frac{1}{(j\omega_{1}+k\omega_{2})^{4}}\quad\mbox{and}
\quad 
  g_{3}(\omega_{1},\omega_{2})=
140{\sum_{j,k}^{'}}\frac{1}{(j\omega_{1}+k\omega_{2})^{6}}.
\end{equation}
Let \(e_{1},e_{2}\) and \(e_{3}\) be the roots of 
\(4z^{3}-g_{2}z-g_{3}=0\); that is,
\begin{equation}
  e_{1}+e_{2}+e_{3}=0,\quad 
e_{1}e_{2}+e_{2}e_{3}+e_{3}e_{1}=-\frac{g_{2}}{4},\quad 
e_{1}e_{2}e_{3}=\frac{g_{3}}{4}. 
\end{equation}
The {\it discriminant} \(\Delta\) of the function \(\wp(u)\) is defined 
by \(\Delta=g_{2}^{3}-27g_{3}^{2}\). If \(\Delta>0\), all roots 
\(e_{1},e_{2}\) and \(e_{3}\) of the equation \(4z^{3}-g_{2}z-g_{3}=0\) 
are {\it real}. Thus, we can assume that \(e_{1}>e_{2}>e_{3}\). 
In this case, 
\(\omega_{1}\) and \(\omega_{2}\) are  given by the formula 
\begin{equation}
\label{eq:omegaint}
\omega_{1}=\int_{e_{1}}^{\infty}\frac{dz}{\sqrt{4z^{3}-g_{2}z-g_{3}}},
\quad 
\omega_{2}=i\int_{-\infty}^{e_{3}}\frac{dz}{\sqrt{g_{3}+g_{2}z-4z^{3}}}.
\end{equation} 
Using a 
rational transformation of a variable as 
   \(v=-\frac{(1-l)(1-m)}{y-\frac{2l+2m-3lm-1}{3}}+1\),
we can rewrite \(s^{-1}(x)\) 
as 
\begin{equation}
s^{-1}(x)=
\int_{\frac{2-l-m}{3}}^{\frac{2l+2m-3lm-1}{3}+\frac{(1-l)(1-m)}{1-x^{2}}}
     \frac{dy}{\sqrt{4y^{3}-g_{2}y-g_{3}}},
\end{equation}
where
 \(g_{2}=\frac{4(1-l+l^{2}-m+m^{2}-lm)}{3}, 
g_{3}=\frac{4(2-l-m)(2l-m-1)(2m-l-1)}{27}\).
We note here that \(4y^{3}-g_{2}y-g_{3}\) can be factored as 
\begin{equation}
  4y^{3}-g_{2}y-g_{3}=4(y-\frac{2-l-m}{3})(y-\frac{2l-m-1}{3})
(y-\frac{2m-l-1}{3}).
\end{equation} 
Thus, we  set
\( e_{1}=\frac{2-l-m}{3}>e_{2}=\frac{2l-m-1}{3}>e_{3}=\frac{2m-l-1}{3}\).
 Thus, 
\(s(x)\) can be written explicitly 
in terms of the Weierstrass elliptic function as
\begin{equation}
\label{eq:rational}
 s^{2}(x)=1-\frac{(1-l)(1-m)}{\wp(\omega_{1}-x)-\frac{2l+2m-3lm-1}{3}}
=\frac{sn^{2}(\sqrt{1-m}x;\sqrt{\frac{l-m}{1-m}})}
 {1-m+msn^{2}(\sqrt{1-m}x;\sqrt{\frac{l-m}{1-m}})}. 
\end{equation}
The function  \(s^{2}(x)\) also has  the same periods  
\(2\omega_{1}\) and 
\(2\omega_{2}\) of the Weierstrass elliptic function \(\wp(x)\) and 
the real period \(2\omega_{1}\) is given by the formula
\begin{equation}
  2K(l,m)=2\omega_{1}=\frac{2K(\frac{l-m}{1-m})}{\sqrt{1-m}}.
\end{equation}
Using
the addition formula of Jacobi \(sn\) function,
 we finally  obtain the {\it explicit} duplication formula of \(s(x)\) as 
\begin{equation}
\label{eq:duplication}
s^{2}(2x)=\frac{4s^{2}(x)(1-s^{2}(x))(1-ls^{2}(x))(1-ms^{2}(x))}
{1+As^{4}(x)+Bs^{6}(x)+Cs^{8}(x)},
\end{equation}
 where 
\(A=-2(l+m+lm),
 B=8 lm\) and 
\(C= l^{2}+m^{2}-2lm-2l^{2}m-2lm^{2}+l^{2}m^{2}\). Thus,
 we obtain the two-parameter family of generalized Ulam-von Neumann maps
\(X_{n+1}=f^{(2)}_{l,m}(X_{n})\)(\ref{eq:formula1}).\\ 
In the same way, we can construct 
{\it generalized cubic maps} \(f^{(3)}_{l,m}\) 
 from the triplication formula 
\(s^{2}(3x)=f^{(3)}_{l,m}(s^{2}(x))\) as
\begin{equation}
\label{eq:gcubic}
Y=f^{(3)}_{l,m}(X)=
\frac{X(-3+4X+\sum_{i=1}^{4}A_{i}X^{i})^{2}}
{1+\sum_{i=2}^{9}B_{i}X^{i}},
\end{equation}
 where \(A_{1},\cdots,A_{4}\) and \(B_{2},\cdots,B_{9}\) 
are given\cite{ku3} by 
\begin{eqnarray} 
\begin{array}{l}
A_{1}=4(l+m),A_{2}=-6(l+m+lm),
A_{3}=12lm,\\
A_{4}=l^{2}+m^{2}-2lm-2l^{2}m-2lm^{2}+l^{2}m^{2},\\
B_{2}=-12(l+m+lm),\\
B_{3}=8(l+m+l^{2}+m^{2}+l^{2}m+lm^{2}+15lm),\\
B_{4}=6(5l^{2}+5m^{2}-26lm-26l^{2}m-26lm^{2}+5l^{2}m^{2}),\\
B_{5}=24(-2l^{2}-2m^{2}-2l^{3}-2m^{3}+4lm+7l^{2}m+7lm^{2})\\
   +24(4l^{3}m+4lm^{3}+7l^{2}m^{2}-2l^{3}m^{2}-2l^{2}m^{3}),\\
B_{6}=4(4l^{2}+4m^{2}+4l^{4}+4m^{4}+17l^{3}+17m^{3}-8lm)\\
     +4(-17l^{2}m-17lm^{2}-17l^{3}m-17lm^{3}-8l^{4}m-8lm^{4})\\
     +4(4l^{2}m^{4}+4l^{4}m^{2}-17l^{3}m^{2}-17l^{2}m^{3}+17l^{3}m^{3}-
54l^{2}m^{2}),\\
B_{7}=24(-l^{3}-m^{3}-l^{4}-m^{4}+l^{2}m+lm^{2}-l^{3}m-lm^{3})\\
     +24(l^{4}m+lm^{4}+4l^{2}m^{2}+4l^{3}m^{2}+4l^{2}m^{3})\\
     +24(l^{4}m^{2}+l^{2}m^{4}-l^{3}m^{3}-l^{4}m^{3}-l^{3}m^{4}),\\
B_{8}=3(3l^{4}+3m^{4}+4l^{3}m+4lm^{3}+4l^{4}m+4lm^{4}-14l^{2}m^{2})\\
  +3(-4l^{3}m^{2}-4l^{2}m^{3}-4l^{3}m^{3}-14l^{4}m^{2}-14l^{2}m^{4}+
4l^{4}m^{3}+4l^{3}m^{4}+3l^{4}m^{4}
),\\
B_{9}=8(-l^{4}m-lm^{4}+l^{3}m^{2}+l^{2}m^{3}+l^{4}m^{2}+l^{2}m^{4}
      -2l^{3}m^{3}+l^{4}m^{3}+l^{3}m^{4}-l^{4}m^{4}).
\end{array}\nonumber
\end{eqnarray}

The generalized cubic map \(f^{(3)}_{l,m}\) has 
the same invariant measure (\ref{eq:density_lm}) as the one of 
the generalized Ulam-von Neumann map.
Furthermore, since 
\begin{eqnarray}
  s^{2}(p_{1}p_{2}x)=f^{(p_{1})}_{l,m}(s^{2}(p_{2}x))
   =f^{(p_{1})}_{l,m}\circ f^{(p_{2})}_{l,m}(s^{2}(x))\nonumber\\
   =f^{(p_{2})}_{l,m}(s^{2}(p_{1}x))
   =f^{(p_{2})}_{l,m}\circ f^{(p_{1})}_{l,m}(s^{2}(x))
   =f^{(p_{1}p_{2})}_{l,m}(s^{2}(x)),\nonumber
\end{eqnarray}
we have the following commutative relations
\begin{equation}
\label{eq:functional}
   f^{(p_{1}p_{2})}_{l,m}(X)=f^{(p_{1})}_{l,m}\circ f^{(p_{2})}_{l,m}(X)
   =f^{(p_{2})}_{l,m}\circ f^{(p_{1})}_{l,m}(X),
\end{equation}
where \(p_{1}\) and \(p_{2}\) are positive integers and 
 \(f^{(1)}_{l,m}(X)\equiv X\). Thus, from the functional relations 
(\ref{eq:functional}), all generalized models of exactly solvable chaos 
given by \(Y=f^{(k)}_{l,m}(X)\) for \(k\in \mbox{\boldmath$Z$}^{+}\) 
can be constructed by generalized models of exactly solvable chaos given by 
\(Y=f^{(p)}_{l,m}(X)\) with \(p\) being primes.
For examples, we can compute \(Y=f^{(k)}_{l,m}(X)\) as follows:
\begin{eqnarray}
   f^{(4)}_{l,m}(X)=f^{(2)}_{l,m} \circ f^{(2)}_{l,m}(X)\nonumber\\
   f^{(6)}_{l,m}(X)=f^{(2)}_{l,m}  \circ  f^{(3)}_{l,m}(X)
   =f^{(3)}_{l,m}  \circ f^{(2)}_{l,m}(X)\nonumber\\
    f^{(8)}_{l,m}(X)=f^{(2)}_{l,m} \circ f^{(2)}_{l,m} 
  \circ f^{(2)}_{l,m} (X)\nonumber\\
   f^{(9)}_{l,m}(X)=f^{(3)}_{l,m} \circ f^{(3)}_{l,m}(X)\nonumber\\
   f^{(p_{1}p_{2}\cdots p_{k})}_{l,m}(X)=
f^{(p_{1})}_{l,m}\circ f^{(p_{2})}_{l,m}
   \circ \cdots f^{(p_{k})}_{l,m}(X).
\end{eqnarray}
All of the above examples of exactly solvable chaos 
are {\it rational mappings} on the unit interval \(\mbox{\boldmath$I$}\) and 
they can be summarized in the following table.\\

\begin{tabular}{@{\Vline\ }l@{\Vline\ }l@{\ \Vline}}
\Hline
\multicolumn{2}{@{\Vline\ }c@{\ \Vline}}
{ {\bf Explicit Invariant Densities of Rational Ergodic Mappings}}
\\ \Hline
\multicolumn{1}{@{\Vline\ }c@{\Vline\ }}{{\bf Rational Mappings}} &
\multicolumn{1}{c@{\ \Vline}}{{\bf Algebraic Densities }}\\ \Hline
{\bf Ulam-von Neumann map} & 
   \(\rho(x)=\frac{1}{\pi\sqrt{x(1-x)}}\)\cite{uf}\\
{\bf  Chebyshev maps} & \(\rho(x)=\frac{1}{\pi\sqrt{x(1-x)}}\)\cite{adler}\\
{\bf Katsura-Fukuda map} & 
\(\rho(x)=\frac{1}{K(l)\sqrt{x(1-x)(1-lx)}}\)\cite{ku3}\\
{\bf Generalized Ulam-von Neumann maps} &  
\(\rho(x)=\frac{1}{K(l,m)\sqrt{x(1-x)(1-lx)(1-mx)}}\)\cite{ku3}\\
{\bf Generalized Chebyshev maps} & 
\(\rho(x)=\frac{1}{K(l,m)\sqrt{x(1-x)(1-lx)(1-mx)}}\) \cite{ku3}\\ 
\Hline
\end{tabular}\\[1cm]
\section{Ritt and Weierstrass's theorems}
The commutativity  
\begin{equation}
   f_{p}\circ f_{q}(x)=f_{q}\circ f_{p}(x)=f_{pq}(x),
\end{equation}
for rational ergodic mappings \(\{f_{p}\}\) comes from the fact that 
their mappings are derived from the multiplication 
formulas of a single elliptic function \(s(x)\) as \(s(pu)=f_{p}[s(u)]\). 
  Thus, they have 
exact solutions of the form  \(X_{n}=s(p^{n}\theta)\). Thus, the
commutativity is the key for
one-dimensional ergodic dynamical systems to have exact solutions.

With regard to the class  of rational permutable functions in this way, Ritt
showed in 1923\cite{ritt} that rational functions constructed from the
multiplication formulas of elliptic 
functions such as the above examples \(\{f_{p}\}\) are  
{\it all} of rational commuting self-maps of the projected line except
trivial cases. 
The trivial cases indicate the cases that rational permutable maps
corresponds to
the  multiplication formulas of rational functions of \(e^{z}\), \(\cos(z)\). 
Consequently,  permutable rational maps which are shown to be ergodic in the
previous sections are essentially  {\it all} of rational permutable ergodic
mappings with 
exact solutions of the form \(X_{n}=s(p^{m}\theta)\), with \(s(x)\) being 
a meromorphic periodic function. 
 Furthermore, if we try to construct ergodic dynamical systems with 
exact solutions without using addition theorems of elliptic functions, we
must avoid some lack of analyticity on a map \(f\) due to the 
Weierstrass theorem on functions possessing an algebraic addition theorem as
follows.
The Weierstrass theorem says that any meromorphic function \(s(u)\)
possessing an 
algebraic addition theorem is either  an elliptic function or is of the form
\(R(u)\) or \(R(e^{\lambda u})\), where \(R\) is a rational function
\cite{prasolov}.
Thus, all of  meromorphic and periodic functions \(s(u)\) possessing an 
algebraic addition theorem belong to the class of elliptic functions.
This is the reason why all of our ergodic rational transformations
constructed here 
does not have a density function in a form
\(\rho(x)=\frac{1}{\sqrt{b_{0}+b_{1}x+b_{2}x^{2}+\cdots+b_{m}x^{m}}}\) with
\(m\geq 5\) whose integral defines  the inverse function of a  hyperelliptic
function with genus \(g(\geq 2)\), but has a density function in  a  form 
\(\rho(x)=\frac{1}{\sqrt{b_{0}+b_{1}x+b_{2}x^{2}+b_{3}x^{3}+b_{4}x^{4}}}\),
whose integral defines  the inverse function of an elliptic function with 
genus \(g(=1)\).  

Of course, not  all of ergodic transformations on \(\mbox{\boldmath$I$}\)
are in 
this category as follows: 

{\bf Example 1: Gauss Map (1845)}\\
 The  Gauss map \(G:x \to 1/x-[1/x]\) which maps
\(\mbox{\boldmath$I$}-\{0\}\) onto \(\mbox{\boldmath$I$}\) is known to be
ergodic with respect to 
the absolutely  continuous invariant measure
\(\mu(dx)=\frac{dx}{\mbox{ln}2(1+x)}\). However, the transformation itself
is not a rational function. 

{\bf Example 2: Bool Transformation (1857)}\\
The transformation of Bool  \(B:x \to x-\frac{1}{x}\) which maps 
\(\mbox{\boldmath$R$}-\{0\}\) onto \(\mbox{\boldmath$R$}\) is known to be 
ergodic \cite{bool}. The transformation preserves infinite measures. However, 
the exact solutions \(X_{n}\) do not seem to exist.

{\bf Example 3:}\\
The two onto mappings \(S_{2}:x \to 2x\quad  \mbox{mod}\quad  1\) and 
\(S_{3}:x \to 3x\quad \mbox{mod}\quad 1\) are mutually permutable 
in the sense that \(S_{3}\circ S_{2}(x)=S_{2}\circ S_{3}(x)\). They are 
 ergodic with 
respect to the Lebesgue measure on \(\mbox{\boldmath$I$}\). 
However, the transformations are not rational. 
\clearpage 
{\bf Example 4:}\\
The two transformations \(F_{2}: x\to \frac{1}{2}(x-1/x)\) 
and 
\(F_{3}: x\to \frac{x(x^{2}-3)}{3x^{2}-1}\) of the real line
\(\mbox{\boldmath$R$}\) are 
mutually permutable in the sense that \(F_{3}\circ F_{2}(x)=F_{2}\circ
F_{3}(x)\). 
They are ergodic with respect to a unique invariant measure
\(\mu(dx)=\frac{dx}{\pi(1+x^{2})}\) (Cauchy distribution) which is
absolutely continuous with respect to 
Lebesgue measure \cite{ku34}. These properties comes from the fact that 
the transformations are
equivalent to the multiplication formula of \(\cot(x)\). Thus, these mappings 
have exact solutions \(X_{n}=\cot(2^{n}\theta)\) and
\(X_{n}=\cot(3^{n}\theta)\).
Hence, maps of example 4, although they are defined on the infinite support,
can be considered to belong to the present category of exactly solvable
chaos.   

\section{Classification problem of exactly solvable chaos}
   It is known that any elliptic function \(s(x)\) can be expressed in 
terms of Weierstrassian elliptic functions \(\wp(x)\) and \(\wp'(x)\) 
 with the same periods, the expression being rational in \(\wp(x)\)
and linear in \(\wp'(x)\)\cite{ww}. 
The invariants \(g_{2}(\omega_{1},\omega_{2})\) and 
\(g_{3}(\omega_{1},\omega_{2})\) of 
Weierstrassian elliptic functions \(\wp(x)\) are
not changed under a transformation  
\begin{equation}
   \omega'_{1}=a\omega_{1}+b\omega_{2},\quad \omega'_{2}=c\omega_{1}+
d\omega_{2},
\end{equation}
 where \(a,b,c\) and \(d\) are integers satisfying \(ad-bc=1\). 
An elliptic modular function \(J\) of \(\wp(x)\)  is given by  
\begin{equation}
\label{eq:modular}
J(\tau )=\frac{g_{2}^{3}}{g_{2}^{3}-27g_{3}^{2}},
\end{equation}
where \(\tau=\frac{\omega_{2}}{\omega_{1}}\). Since the elliptic 
modular function \(J(\tau)\) in Eq.(\ref{eq:modular})
 is invariant under a certain condition; namely, 
\begin{equation}
  J(\frac{c+d\tau}{a+b\tau})=J(\tau),\quad ad-bc=1,
\end{equation}
 where \(a,b,c\) and \(d\) are integers,
\(J(\tau)\) of an elliptic function \(s(x)\) can serve as a characteristic 
 of exactly solvable chaos. For an example, let us consider 
the classification problem of the two-parameter family of 
the generalized Ulam-von Neumann maps \(X_{n+1}=f^{(2)}_{l,m}(X_{n})\).
In this case, it is easy to check that 
\begin{equation}
  J(l,m)=\frac{4[(1-m)(1-l)+(l-m)^{2}]^{3}}{27(1-l)^{2}(1-m)^{2}(l-m)^{2}}.
\end{equation} 
Thus,  the equality
   \(J(l,m)=J(l',m')\) gives us the solutions of the classification problem
 as follows:
\begin{equation}
\label{eq:condition}
  \lambda=\lambda',\lambda=1-\lambda',\lambda=\frac{1}{1-\lambda'},
\lambda=\frac{1}{\lambda'},\lambda=\frac{\lambda'}{\lambda'-1},
\lambda=\frac{\lambda'-1}{\lambda'},
\end{equation}
where \(\lambda\equiv \frac{l-m}{1-m}\) and 
\(\lambda'\equiv\frac{l'-m'}{1-m'}\).
This means that when one of 
 the conditions in Eq.(\ref{eq:condition}) is satisfied,
 dynamical systems \(X_{n+1}=f^{(2)}_{l,m}(X_{n})\) and 
\(Y_{n+1}=f^{(2)}_{l',m'}(Y_{n})\) have an algebraic relation
such that the relation \(H[f^{(2)}_{l,m}(X),f^{(2)}_{l',m'}(X)]=0\) 
holds, where  
\(H\) is a polynomial  function in two variables. In this way, we can 
solve the classification problem of  exactly solvable chaos 
\(X_{n+1}=f(X_{n})\).\\
 
\section{Summary and Discussions}
Rational ergodic maps on the unit interval are systematically constructed
from addition theorems of elliptic functions in a unified manner. According
to the classical Weierstrass's theorem about meromorphic functions
possessing algebraic addition theorems and Ritt's theorem about permutable
rational functions, such  rational mappings due to addition theorems of
elliptic functions are essentially 
{\it all} rational ergodic transformations that have exact solutions as well
as exact 
density functions.
Therefore, in discrete-time systems, solvable models showing chaotic behavior 
do not belong to an  exceptional class but they are ubiquitous as the
presence of 
multiplication theorems of 
elliptic functions. This implies the gap between the notion of solvability
and integrability in discrete-time dynamical systems as follows:  The notion
of solvability in discrete-time systems can be 
 compatible with chaotic behavior which are commonly believed to 
indicate non-integrability  of the systems, whereas the notion of the
solvability  is essentially same as that of the integrability
in continuous-time systems (ODE). Thus, 
further studies of asking what is an essential characteristic distinguishing
chaotic discrete-time systems from regular discrete-time systems are
clearly  needed for the settlement of the issue.\\
  
{\bf Acknowledgements}\\
The author would like to acknowledge  K. Iguchi and  A. Bobenko for useful
discussions. This work was, in part, supported by RIKEN-SPRF grant. 


\begin{thebibliography}{99}
\bibitem{ww1} Any elliptic function \(w(u)\)  has an algebraic relation 
 with the Weierstrass elliptic function \(\wp (u)\) with the same 
periods as
  \(G(w,\wp)=0\),   
\(G\) being  a rational function in two variables.
\end{thebibliography}
\end{document}